\documentclass[a4paper,french]{rnti}
\usepackage[T1]{fontenc}
\usepackage[latin1]{inputenc}
\usepackage{subcaption}
\usepackage{tikz}
\usepackage{mathtools}
\usepackage{amsmath,amsfonts,amssymb}

\usepackage{url}

\usepackage{graphicx}

\usetikzlibrary{arrows, positioning, shadows, calc,trees,backgrounds,shapes,decorations.pathmorphing,fit,positioning,chains}
\definecolor{lavander}{cmyk}{0,0.48,0,0}
\definecolor{violet}{cmyk}{0.79,0.88,0,0}
\definecolor{burntorange}{cmyk}{0,0.52,1,0}
\def\lav{gray!50}
\def\oran{orange!50}
\tikzstyle{peers}=[draw,circle,fill=\lav, \lav, minimum size=1pt, thin, align=center, anchor=base, text=black]
\tikzstyle{superpeers}=[draw, minimum height=1.4em, minimum width=0.7em, fill=\oran, double copy shadow={shadow xshift=2pt, shadow yshift=4pt, fill=white, draw}, draw, rectangle, \oran, thin, align=center, anchor=base, text=black]
\tikzstyle{legendsp}=[rectangle, draw, rounded corners, thin,fill=\oran, \oran, minimum width=1.5cm,align=center, anchor=base, text=black]
\tikzstyle{legendp}=[rectangle, draw, rounded corners, thin, fill=\lav, \lav, minimum width= 1.5cm,align=center, anchor=base, text=black]

\newcommand\myeq{\stackrel{\mathclap{\tiny\mbox{rang}}}{=}}

\titrecourt{Peerus Review: Un outil de recherche d'experts scientifiques}

\nomcourt{R. Brochier et al.}

\titre{Peerus Review: Un outil de recherche d'experts scientifiques}

\auteur{Robin Brochier\affil{1}\affilsep\affil{2},
        Adrien Guille\affil{1},
        Julien Velcin\affil{1}\\
        Benjamin Rothan\affil{2},
        Fran\c cois Di Cioccio\affil{2}}
        
\affiliation{
    \affil{1}Laboratoire ERIC, 5 Avenue Pierre Mendès France, 69500 Bron\\
          robin.brochier@univ-lyon2.fr, julien.velcin@univ-lyon2.fr, adrien.guille@univ-lyon2.fr\\
          \http{https://eric.ish-lyon.cnrs.fr/}\\
    \affil{2}DSRT, 103 avenue du Marechal de Saxe 69003 Lyon \\
          robin@peer.us, benjamin@peer.us, francois@peer.us\\
          \http{https://peer.us/}\\          
 }

\resume{%
Nous proposons un outil de recherche d'experts appliqué au monde académique sur les données générées par l'entreprise \textit{DSRT} dans le cadre de son application \textit{Peerus} \footnote{https://peer.us/}. Un utilisateur soumet le titre, le résumé et optionnellement les auteurs et le journal de publication d'un article scientifique et se voit proposer une liste d'experts, potentiels reviewers de l'article soumis. L'algorithme de recherche est un système de votes reposant sur un modèle du langage entrainé à partir d'un ensemble de plusieurs millions d'articles scientifiques. L'outil est accessible à chacun sous la forme d'une application web intitulée \textit{Peerus Review} \footnote{https://review.peer.us/}.     
}

\summary{%
We propose a tool for experts finding applied to academic data generated by the start-up \textit{DSRT} in the context of its application \textit{Peerus}. A user may submit the title, the abstract and otpionnally the authors and the journal of publication of a scientific article and the application then returns a list of experts, potential reviewers of the submitted article. The retrieval algorithme is a voting system based on a language modeling technique trained on several millions of scientific papers.       
}

\begin{document}
%\layout

% DEBUT DE L'ARTICLE
%
\section{Introduction}

L'évaluation par des relecteurs est un processus scientifique par lequel les experts d'une discipline vérifient la qualité du travail de leurs pairs. L'examen et la validation des travaux scientifiques est une pierre angulaire de la recherche scientifique. Le nombre croissant de publications journalières dans un contexte académique de plus en plus compétitif (publier ou périr) et la digitalisation du monde de l'édition légitiment le développement d'outils informatiques d'aide au processus de \textit{reviewing}.

Lorsqu'un chercheur propose un article à un éditeur, ce dernier est en charge de trouver un certain nombre de reviewers. Le temps nécessaire pour trouver ces reviewers constitue le principal goulot d'étranglement de l'édition scientifique, retardant parfois la date de publication d'un article de plusieurs mois. Nous proposons un outil de recherche d'experts scientifiques nommé \textit{Peerus Review} \footnote{Inscription gratuite et démonstration disponible}. Les experts recherchés sont les reviewers potentiels d'un article-requête émis sous la forme d'un titre et d'un résumé. 

L'algorithme de recherche d'experts s'appuie sur les données générées par l'entreprise \textit{DSRT} afin d'estimer la probabilité de chaque scientifique de la base d'être un expert de l'article soumis. Cette estimation est réalisée en deux étapes, la première calculant les similarités entre la requête et les articles de la base de données, puis la seconde agrégeant ces similarités par un système de votes où chaque article renforce le score de ses auteurs.    

\section{Modèle implémenté}

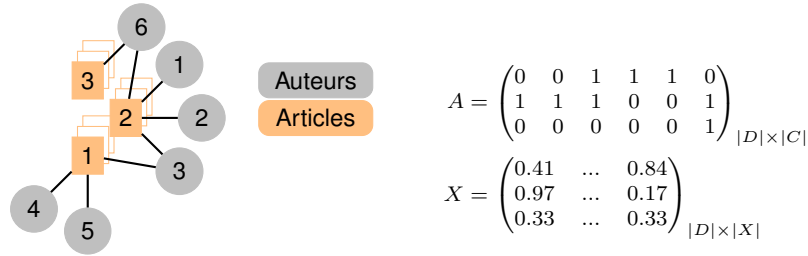
\begin{figure}[t!]
\centering
    \begin{subfigure}[b]{0.4\textwidth}
        \centering
        \begin{tikzpicture}[font=\sffamily\small, auto, thick, scale=0.25]
          \edef\mya{0}
          \foreach \place/\name in {{(0,-2)/a}, {(2,0)/b}, {(0,2)/d}}
           \pgfmathparse{int(\mya+1}
            \xdef\mya{\pgfmathresult}
            \node[superpeers] (\name) at \place {\mya};
           \foreach \pos/\i in {above right of/1, right of/2, below right of/3}
            \node[peers, \pos =b ] (b\i) {\i};
           \foreach \speer/\peer in {b/b1,b/b2,b/b3}
           \path (\speer) edge[-] (\peer);
           \path (a) edge[-] (b3);
           \node[peers, above right of=d] (d1){6};
           \path (d) edge[-] (d1);
           \path (b) edge[-] (d1);
           \edef\mya{3}
           \foreach \pos/\i in {below left of/1, below of/2}
            \pgfmathparse{int(\i+3)}
            \edef\mya{\pgfmathresult}
            \node[peers, \pos =a ] (a\i) {\mya};
           \foreach \speer/\peer in {a/a1,a/a2}
           \path (\speer) edge[-] (\peer);
           \node[legendsp] at (12,0) {\small{Articles}};
           \node[legendp] at (12,2) {\small{Auteurs}};
        \end{tikzpicture}
        \caption{Graphe biparti de la base de données où l'on liste les auteurs et articles.}
\end{subfigure}%
~~~~
\begin{subfigure}[b]{0.4\textwidth}
        \centering \footnotesize
        \begin{align*}
            A &= \begin{pmatrix}
              0        & 0       & 1       & 1         & 1        & 0\\
              1        & 1       & 1       & 0         & 0        & 1\\
              0        & 0       & 0       & 0         & 0        & 1\\
            \end{pmatrix}_{|D|\times|C|}\\
            X &= \begin{pmatrix}
              0.41        & ...   & 0.84  \\
              0.97        & ...   & 0.17  \\
              0.33        & ...   & 0.33  \\
            \end{pmatrix}_{|D|\times|X|}%\\
            %q &= (0.14, ..., 0.96)_{|X|}
        \end{align*}
        \caption{Représentation matricielle correspondante.}
   \end{subfigure}
   \caption{Représentation matricielle d'un jeu de données pour la recherche d'experts.}
   \label{fig:data}
\end{figure}

On considère un graphe biparti $G=(V,E)$ composé de deux types de n\oe{}uds $V = V_C \cup V_D$ correspondant aux $C$ auteurs et $D$ articles, dont les liens $E$ sont les associations auteur-article. On note $A$ la matrice d'adjacence de $G$. On dispose en plus d'une matrice d'attributs $X$ pour l'ensemble des articles.  La recherche d'experts consiste à générer, étant donné un ensemble $(A,X)$ (voir figure \ref{fig:data}) et une requête $q$ composée de texte, les probabilités (ou scores) $s=(s_1,s_2,...,s_{|C|})$ des auteurs $c=(c_1,c_2,...,c_{|C|})$ d'être experts de cette dernière.

En pratique, la requête est la concaténation du titre et du résumé et est transformée en vecteur par une technique de modélisation du langage. Nous avons testé différentes techniques telles que \textit{TF-IDF}, \textit{LSA}, \textit{LDA} et \textit{word2vec}. Cette dernière technique est actuellement utilisée et nécessite plus de traitements puisqu'elle ne fournit pas directement de représentation vectorielle pour les articles, mais seulement pour les mots qui les composent. Deux étapes sont ensuite nécessaires pour estimer les scores des auteurs:

\begin{itemize}
\item \textbf{similarité requête-documents}: la proximité entre la requête et les articles de la base est estimée. Pour ce faire, on calcule la similarité cosinus pour chaque pair de représentations vectorielles requête-article. Pour \textit{word2vec}, la distance entre deux articles est calculée grâce à la distance du cantonnier (ou métrique de Wasserstein) décrite dans \cite{kusner2015word} à partir des représentations lexicales de leurs mots \footnote{Cette distance s'inspire du problème du transport et consiste à trouver la distance minimum à parcourir pour se déplacer de l'ensemble des vecteurs du premier article vers l'ensemble du second.}.     
\item \textbf{associations documents-auteurs}: on utilise une technique de fusion de données pour attribuer un score à chaque auteur. Celle utilisée dans notre prototype est nommée rang-réciproque et consiste dans un premier lieu à classer les articles selon l'ordre décroissant de leurs similarité à la requête, puis d'agréger les rangs $rank_d(q)$ des articles écrits par chaque auteur selon la formule $RR_{auteur} = \sum\limits_{d \in D(e)} \frac{1}{rank_d(q)}$ où $D(e)$ est l'ensemble des articles écrits par l'auteur $e$ (correspondant à la ligne $e$ de la matrice $A$) \footnote{On réalise de la sorte un compromis entre la similarité sémantique des articles à la requête et le nombre de publications de chaque auteur afin d'estimer l'expertise de ceux-ci.}. 
\end{itemize}

Ce modèle de votes est l'un des algorithmes confrontés dans \cite{macdonald2006voting}. Dans \cite{balog2012expertise}, les systèmes de votes sont vus comme une altération du second modèle génératif présenté dans \cite{balog2006formal} où la probabilité d'un auteur d'être un expert étant donnée une requête est estimée en utilisant la formule de Bayes: $P(e|q) \myeq P(q|e)P(e) = \sum\limits_{d \in D(e)} P(q|d)P(d|e)P(e)$, $P(q|d)$ étant la similarité entre l'article $d$ et la requête $q$, $P(d|e)$ étant la force d'association entre l'article $d$ et l'auteur $e$ et $P(e)$ est la probabilité \textit{a priori} de l'auteur $e$ d'être un expert (souvent considérée comme uniforme).   

\section{Interface utilisateur}

\begin{figure}[t!]
\center
 \includegraphics[width=10cm]{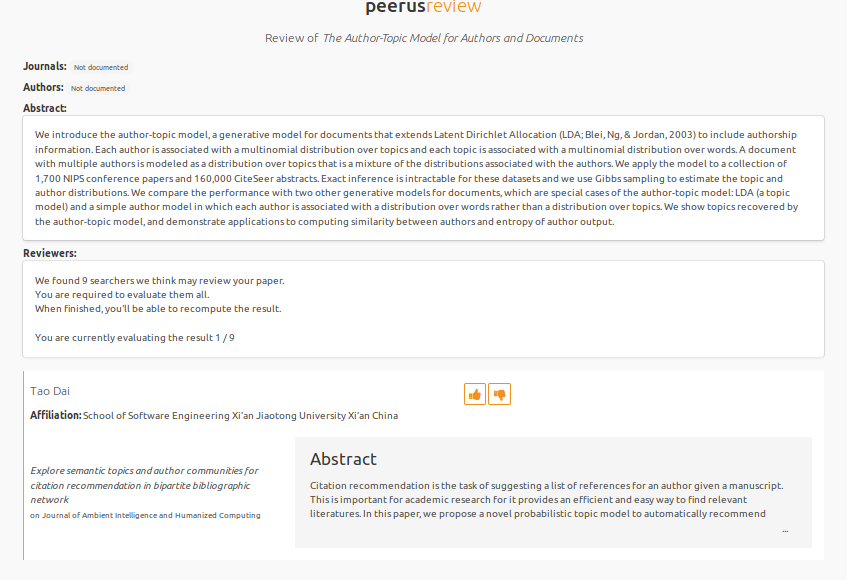}
   \caption{Présentation du profil d'un auteur, potentiel reviewer, retourné par \textit{Peerus Review}. L'utilisateur peut donner son avis sur la pertinence de l'expert proposé.}
   \label{fig:img}
\end{figure}

L'utilisateur de \textit{Peerus Review} soumet une requête à travers un formulaire HTML constitué de deux entrées textuelles pour le titre et le résumé. L'algorithme procède alors au classement et retourne la liste des neuf meilleurs auteurs. Les résultats sont présentés séquentiellement (voir figure \ref{fig:img}), à raison d'une page par auteur, afin de garantir deux objectifs:
\begin{itemize}
\item l'utilisateur doit pouvoir vérifier, par lui-même, la véracité des résultats. Pour cela, l'intégralité des informations liées aux articles des auteurs est présentée (titre, résumé, noms d'auteurs, affiliation et date de publication). 
\item \textit{Peerus Review} doit pouvoir collecter la satisfaction des utilisateurs vis-à-vis des résultats retournés. C'est pourquoi l'utilisateur doit valider ou invalider un auteur qui lui est présenté afin de pouvoir consulter le suivant. Cela permet d'estimer la qualité de l'algorithme et l'utilisateur peut requérir un classement moyennant le vecteur requête avec les vecteurs des articles qu'il estime satisfaisants (option <<recompute>>). 
\end{itemize}

\section{Travaux futurs et conclusion}
Notre prototype permet de rechercher des reviewers potentiels à un article soumis grâce à un système de votes reposant sur une technique de modélisation du langage indépendamment sélectionnée. Dans des travaux futurs, nous exploiterons la topologie des relations auteurs-articles à travers des modèles de propagation dans les graphes (\cite{serdyukov2008modeling}). \`A plus long terme, nous explorerons les techniques d'apprentissage de représentations plongeant dans un même espace experts et articles scientifiques (\cite{van2016unsupervised}).

\bibliographystyle{rnti}
\bibliography{biblio}

\Fr

\end{document}